# DARHT Axis 1 Compensation Can D-dot Calibration

J. Martin Taccetti, Darin Westley, Jeffrey B. Johnson
*Los Alamos National Laboratory, P.O. Box 1663*
*Los Alamos, NM 87545 USA*

Paul Flores
*National Security Technologies, LLC*
*Los Alamos, NM 87544 USA*

## Abstract

We detail the calibration of the D-dot probes used on the Los Alamos National Laboratory DARHT Axis 1 LIA compensation cans (CCs). We hope this will serve not only as a record of this calibration, but also as a guide on how to perform similar calibrations on other systems where D-dots are deployed. Although a simple measurement, its difficulty lies in the fact that the same geometry used when fielded must be used during calibration, the ultimate goal being to measure the capacitance between the probe sensor and the component whose voltage one intends to measure. Because of its linearity with voltage, this can be done at a lower voltage than during fielding, the only caveat being that this lower voltage pulse must still provide enough signal to noise. After a brief summary of our results, we include a description of a compensation can, the D-dot probe and its operation. We then provide 3D simulation results of the capacitance of this probe to the high voltage bushing under test inside of the compensation can. Finally, we describe our calibration setup and discuss our results.

## A.  Introduction/Summary of results

We calibrated three different D-dot probes to obtain an average calibration, since all of the D-dots used on the DARHT Axis 1 LIA and the CCs they are fielded on are of identical construction. (Although we use the term *D-dot* to refer to these capacitive voltage probes, they are also referred to as *E-dots* or *V-dots* in the literature.) The measured scaling factors and capacitances for each of these D-dots are shown in Table 1. The average values are **(4.949 ± 0.023) × 10$^5$** and **(101.8 ± 0.48) fF**, respectively. Details on how these values were obtained and how they are applied in the data acquisition system are provided in a later section.

## B.  Description of the Compensation Can

A cross-section of a CC is shown in Fig. 1. The input cable plugs into the top. The HV pulse is routed via the *conductive disk* to the LIA gap via the *drive rod*, with a parallel compensation resistor path to ground. The entire volume is filled with insulating oil (Shell DIALA-AX). The D-dot probe (not shown) is located in the D-dot receptacle on the figure, with its pick-up button facing the conductive disk. (The piece holding the conductive disk to the CC housing is a Rexolite insulator.)

Table 1. Scaling factors for Axis-1 Compensation Can D-dots.

| D-dot SN | Cal Scale Factor $\phi \pm \sigma$ (Gauss Fit) | | Capacitance $\pm \sigma$ (×10$^{-15}$ F) | | Integrator RC ($\mu$s) |
|---|---|---|---|---|---|
| 1E406 | 4.975×10$^5$ | ±0.018×10$^5$ | 101.3 | ± 0.37 | 1.26 |
| 1E107 | 4.943×10$^5$ | ±0.010×10$^5$ | 102.0 | ± 0.21 | 1.26 |
| 1E395 | 4.929×10$^5$ | ± 0.014×10$^5$ | 102.3 | ± 0.29 | 1.26 |
| **Average** | **4.949×10$^5$** | **± 0.023×10$^5$** | **101.8** | **± 0.48** | |

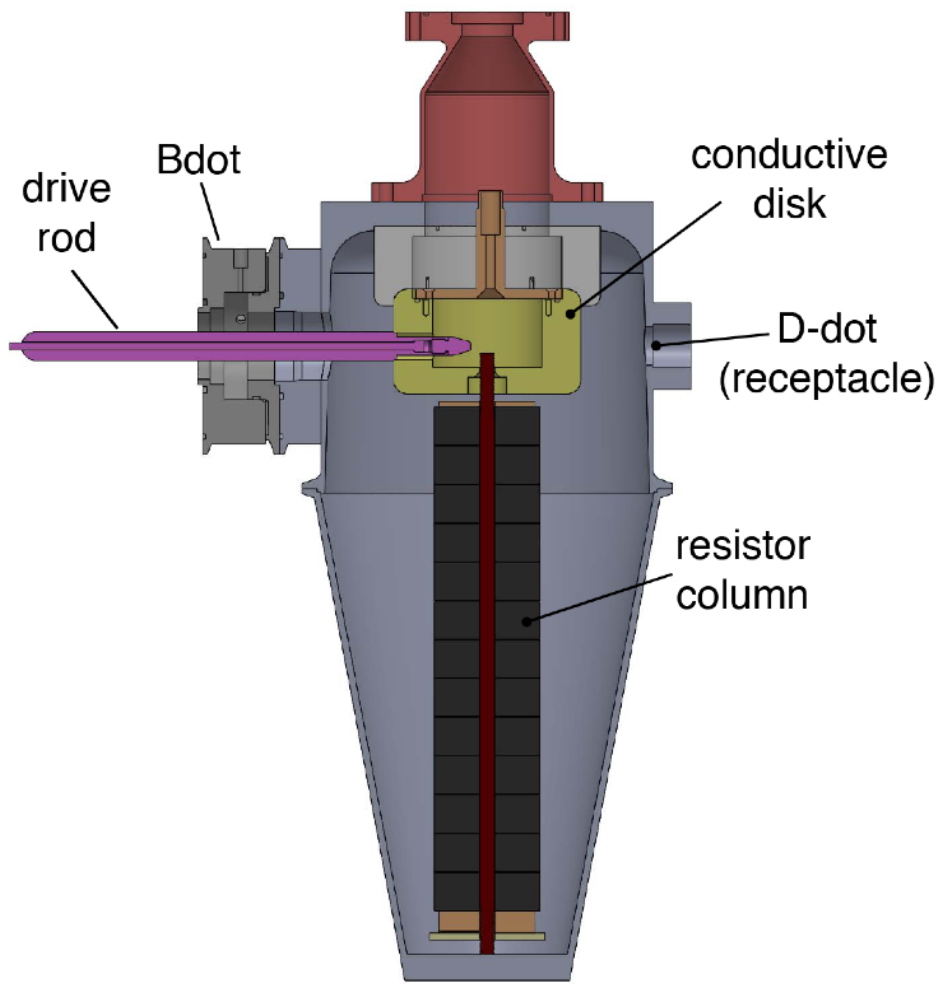

Figure 1. Compensation can cross-section.

### C. Description of the D-dot probe

A diagram of the D-dot probe is shown in Fig. 2. It consists of a pick-up 'button' connected to the center pin of an 'N'-type connector, electrically isolated from the ground outer sleeve by a thin Rexolite sleeve, shown on the figure in yellow (pick-up button is labeled 5, connector is labeled 7). Operation of this type of probe was described by Ekdahl (1). The D-dot, or capacitive voltage sensor circuit is basically a capacitive voltage divider, and is reproduced from (1) in Fig. 3. In this circuit, $C_1$ represents the capacitance between the sensor button and the object to be measured, while $C_2$ represents the capacitance between the button and ground. $R_1$ represents a series resistance (which in our case is negligible), and $R_2$ represents the cable termination resistance, 50 $\Omega$. The sensor mode of operation depends on the value of $RC =$ $(R_1+R_2) \times (C_1+C_2)$ relative to the fastest rate of change $\tau$ of the signal to be measured. As long as $RC \ll \tau$, the output signal $V_2$ is proportional to the *derivative* of the voltage to be measured, $V_1$.

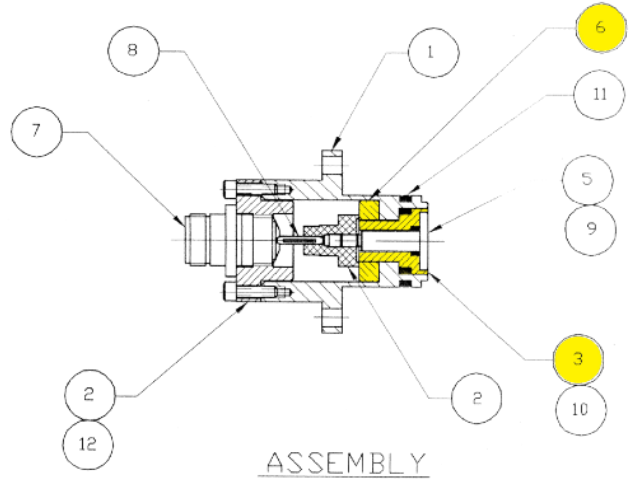

Figure 2. D-dot cross-section, with Rexolite insulating parts shown in yellow.

$$V_2 = R_2 C_1 \frac{dV_1}{dt} \tag{1}$$

Capacitance $C_2$, from sensor button to ground, can be approximated (ignoring fringe fields) using the dimensions of the probe shown in Fig. 2, including only the range spanned by the Rexolite parts. This results in $C_2 \approx 5.3$ pF. Capacitance $C_1$, from the button to the CC conductive disk, is more complicated to estimate due to the 3-dimensional field structure. A full 3-D calculation was performed, and is described in the next section. A quick estimate, to at least provide the order of magnitude expected, can be performed by assuming a coaxial structure with the inner diameter equal to the diameter of the conductive disk, with a length equal to the diameter of the sensor button. Using this approximation, and multiplying the resulting capacitance by the ratio of the outer circumference to the probe diameter, we obtain that $C_1 \approx 107$ fF. With negligible $R_1$, this means our $(R_1+R_2) \times (C_1+C_2) = 0.27$ ns, and we are clearly in the regime where Eq. (1) is valid, since our drive pulse risetime is $\leq 20$ ns. The relative permittivities of the materials used for these calculations, as well as on the 3-D simulations below, are shown on Table 2.

From Eq. (1), and after hardware integrating the signal $V_o$ with (with integrator time constant $RC$), into a scope channel also terminated in $R_L$ ($R_L \equiv R_2$ in Eq. (1)), the recorded signal

$$V_{recorded} = \frac{1}{2RC} \int V_{out}\, dt = \left(\frac{R_L C_1}{2RC}\right) V_1, \tag{2}$$

so that the signal at the D-dot is

$$V_1 = \left(\frac{2RC}{R_L C_1}\right) V_{recorded} \equiv \phi V_{recorded}, \tag{3}$$

where we define

$$\phi = \frac{2RC}{R_L C_1} \tag{4}$$

as the *scale factor* to be entered into our data acquisition system. The capacitance $C_1$ is calculated from Eq. (4). Using error propagation, the uncertainty for $C_1$ is given by

$$\sigma_{C_1} = \sigma_\phi \left( \frac{2RC}{R_L \phi^2} \right), \quad (5)$$

where $\sigma_\phi$ is the uncertainty in the scale factor as obtained from our measurements.

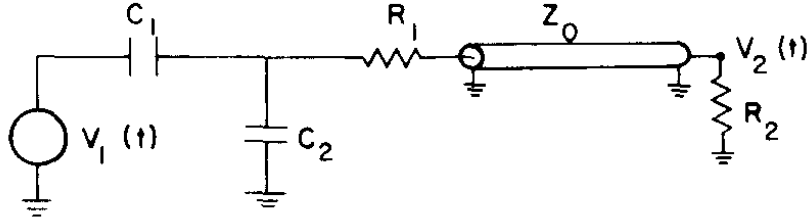

Figure 3. Equivalent D-dot circuit. Reproduced from Ekdahl, C. A., Rev. Sci. Instrum. **51**(12), 1645 (1980), with the permission of AIP Publishing.

If the calibration scaling factor $\phi$ was obtained with an integrator with time constant $\tau_{cal}$, and the integrator used during recording has a time constant $\tau_{rec}$, we see from Eq. (4) that the scaling factor to use is $\phi_{rec} = \phi_{cal}(\tau_{rec}/\tau_{cal})$.

### D. 3-D Simulation of CC conductive disk region

Due to the 3-D structure of the fields detected by the D-dot inside the CC, a 3-D simulation was used to calculate the capacitance between the sensor button and ground. We simulated this on ANSYS Maxwell 3D.

The model includes the top 'half' of the CC that encloses the conductive disk. A cut-away view of the setup is shown in Figs. 4 and 5. The D-dot sensor button is modeled as a simple cylinder, so that only the sensor to electrode capacitance is calculated with this simulation. The same material parameters presented on Table 2 were used on these simulations. We calculated $C_1$ = 103.5 fF. These results are consistent with the result obtained from our quick estimate in Sec. C.

Figure 6 shows the equipotentials on planes perpendicular to the D-dot probe surface. The top of the figure shows a vertical cut along the mid-plane of the probe button, while the bottom shows a horizontal cut.

### E. Calibration setup

In order to perform this calibration, we came up with a setup that includes only a subset of the CC components, and only those that would affect the measurement. The main design considerations of the assembly were to 1) maintain electrical conductivity through the surface of the test stand, and 2) to electrically isolate the center conductor of the assembly from this conducting surface. The former was accomplished by surface depositing a chromate chemical film (MIL-C-5541 Type II), and the latter by fabricating a cylindrical High Density Polyethylene (HDPE) insulator to hold the conducting center disk in place. The completed assembly consisted of the newly designed parts assembled with the existing top section of the CC, which was filled with insulating oil (Fig. 7). Essentially, the top housing (Fig. 8), cable input, Rexolite insulator and conductive disk are unchanged. A new bottom cover with an oil connector was made and an HDPE spacer keeps the conductive disk tight to this cover. Another cover on the drive rod side was made with an 'N'-type connector whose inner conductor is connected to the conductive disk. (Both covers are made to use Voss Ind. V-retainer clamps: VOSS#8125A-575-A on small side cover, VOSS#8125A-975-A on large bottom cover - refer to fig 5a. in (2). The interface angle referred to in that figure should be 40 degrees, as also shown in Fig. 8.) A stand was also constructed to solidly attach the entire assembly to a table. Note that the as-built dimensions of the top housing (Fig. 8) were slightly different than the schematic - we measured the distance from the probe button to the conductive disk and found it to be (1.718 ± 0.003)", while the schematic indicates 1.789". The measured value was used in the simulations.

Table 2. Relative permittivities of materials used in calculations.

| Material | $\epsilon_r$ |
|---|---|
| Shell DIALA-AX oil | 2.25 |
| Rexolite | 2.53 |

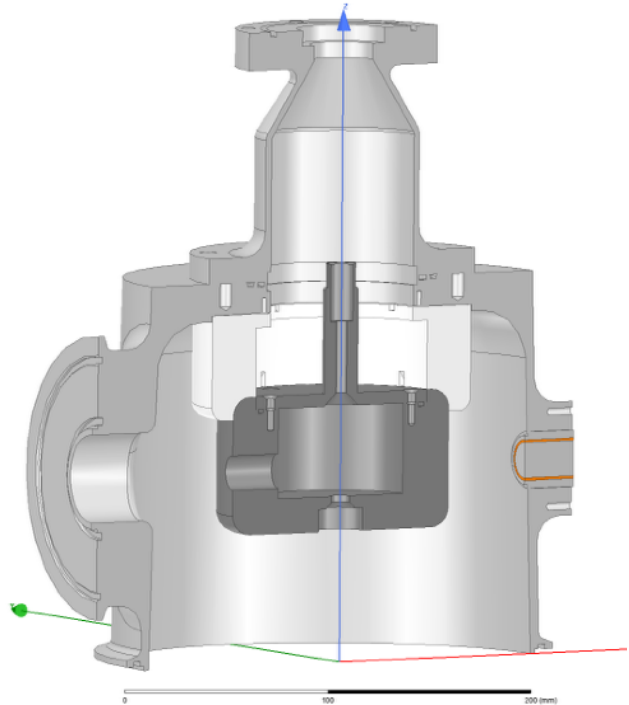

Figure 4. ANSYS Maxwell 3D problem setup. The D-dot probe button is modeled as a solid cylinder.

An input pulse, with risetime and pulse length similar to a pulse typically used to drive the cell, although only 7 kV in amplitude, is input via the top cable connection through the conductive disk. A matched 50 ohm high-voltage Barth load is attached to conductive disk via the 'N'-type connector. The signals measured on the Barth and on the D-dot are compared (see Sec. F for details) to obtain the scale factor.

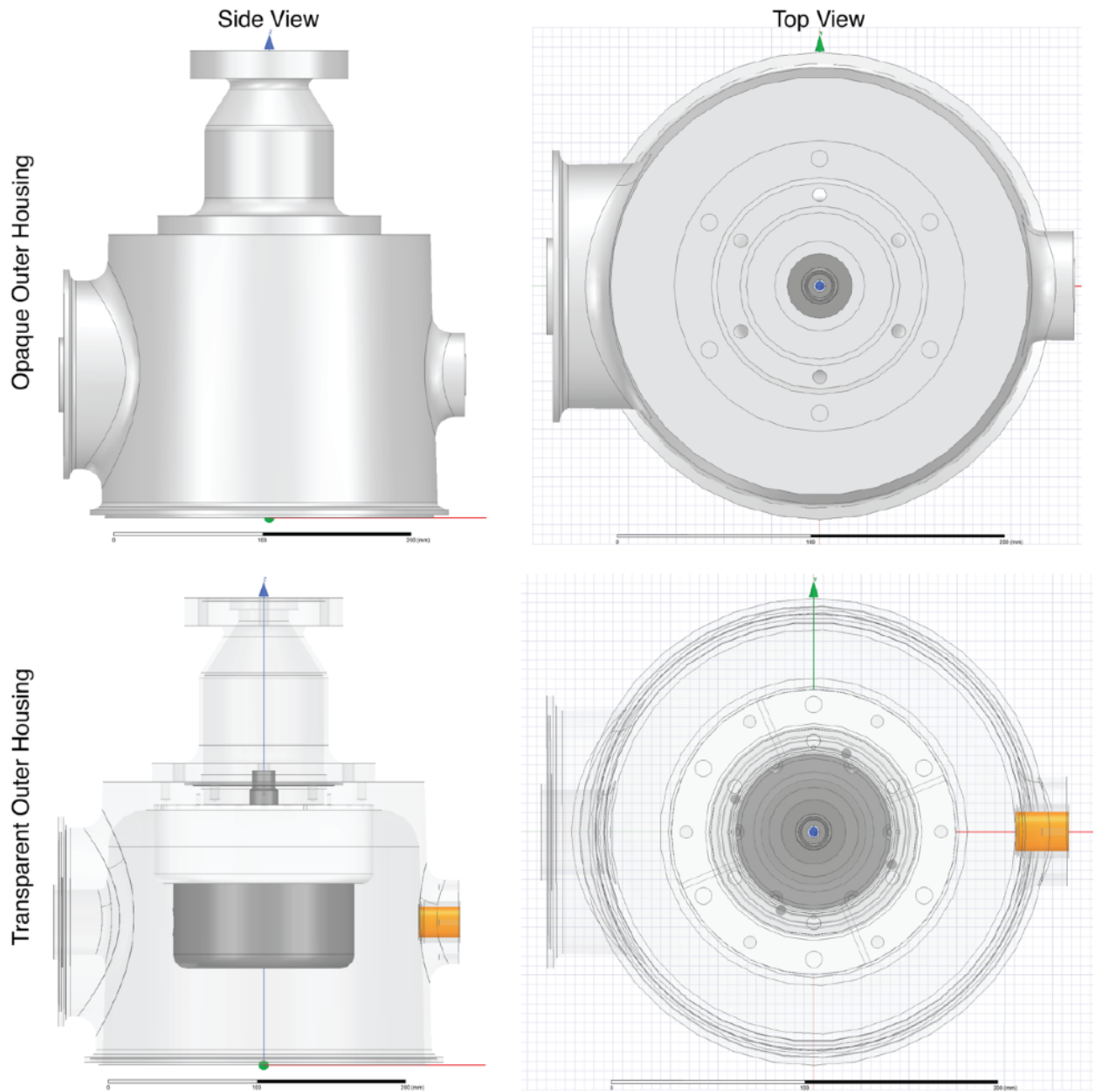

Figure 5.   Additional views of the ANSYS Maxwell 3D problem setup.

**F.   Measurements and description of analysis.**

The recording setup is shown in Fig. 9. Our data acquisition system was utilized for analyzing the data collected for these calibrations. All of the Barth load reference data was baseline corrected and multiplied by the associated attenuation factor (voltage ratio of 1989:1). The D-dot data was clipped, baseline corrected, time shifted, and droop corrected (with the relevant RC constant). The reference pulse (Barth data) was divided by the corrected D-dot data, and the calibration factor was determined by taking the mean of all y-values in a 50 ns window roughly in the center of the peak of the pulse. These values were then applied to the reference pulse and overlaid with the corrected D-dot data for confirmation.

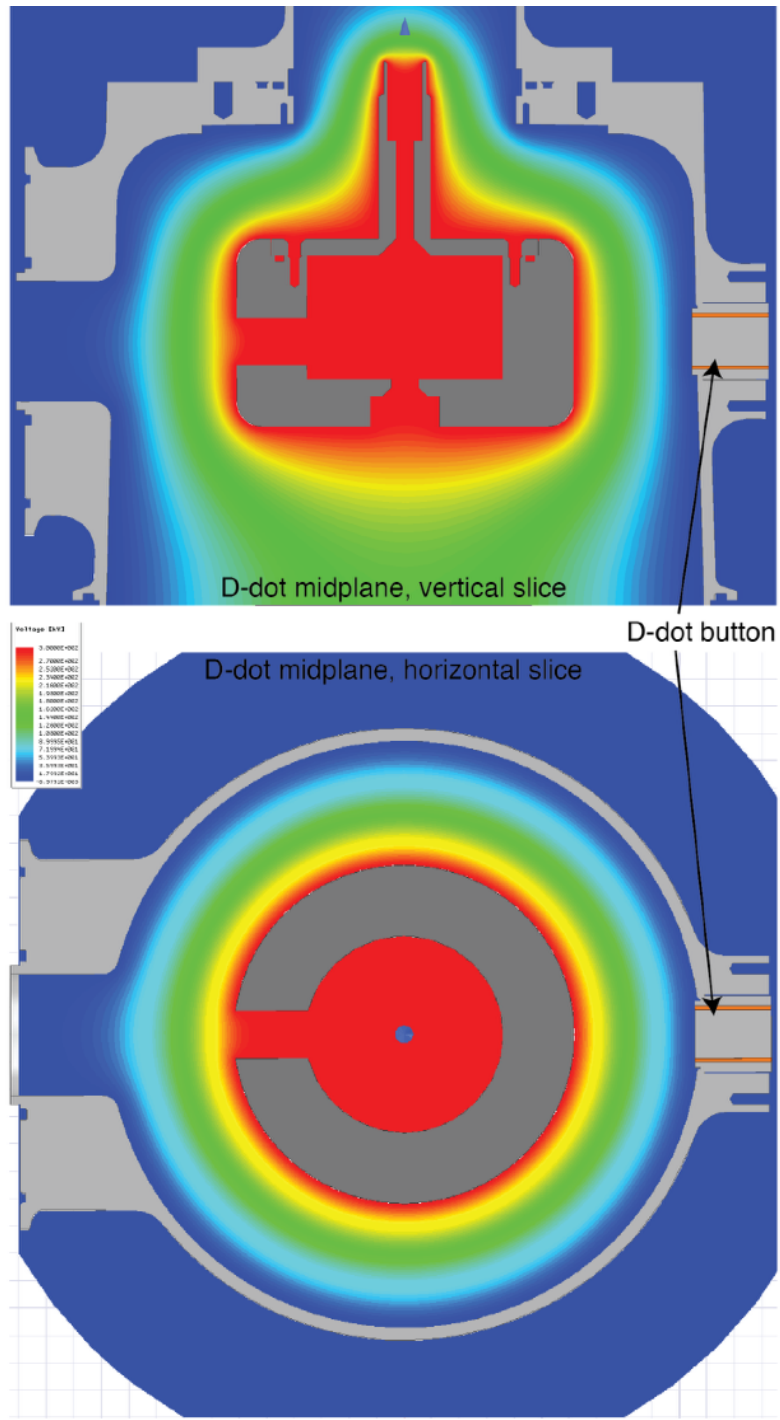

Figure 6. Equipotentials on two D-dot mid-planes: top is a vertical slice; bottom is a horizontal slice.

The resulting scaling factors for each of the three D-dots tested are summarized in Tables 3-5. The average for each is also listed. The corresponding waveforms showing the analyzed D-dot data overlaid with the corresponding Barth data are shown in Figs. 10 - 15.

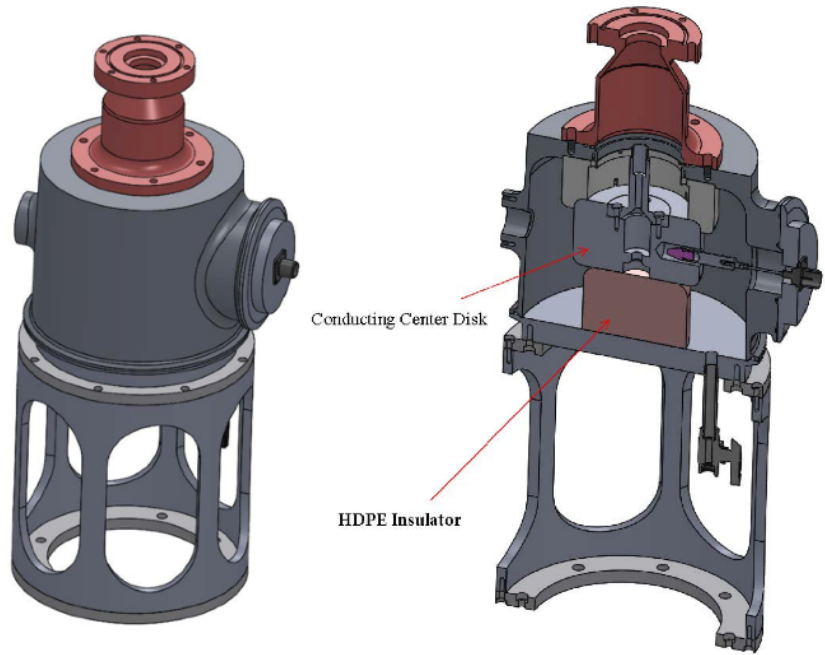

Figure 7.   Compensation can D-dot calibration assembly.

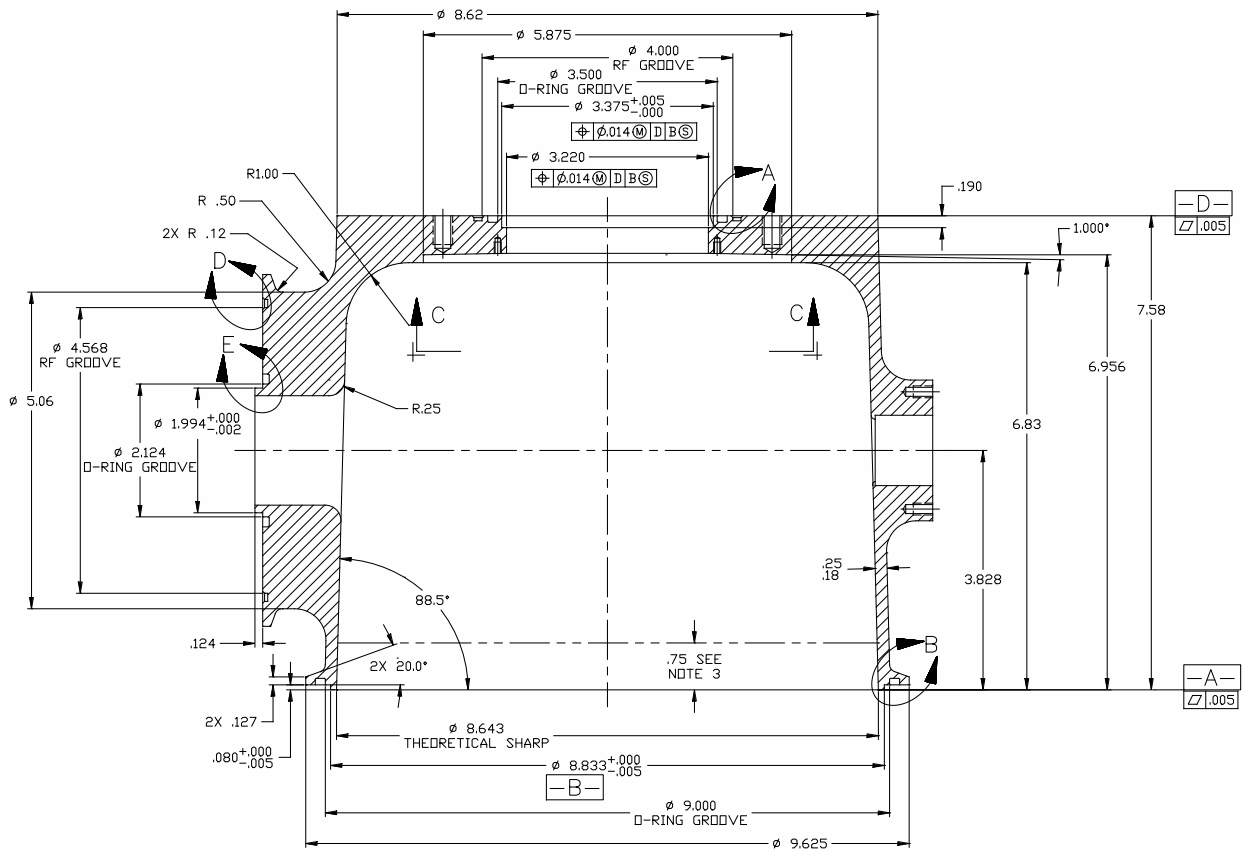

Figure 8.   Schematic of CC top housing.

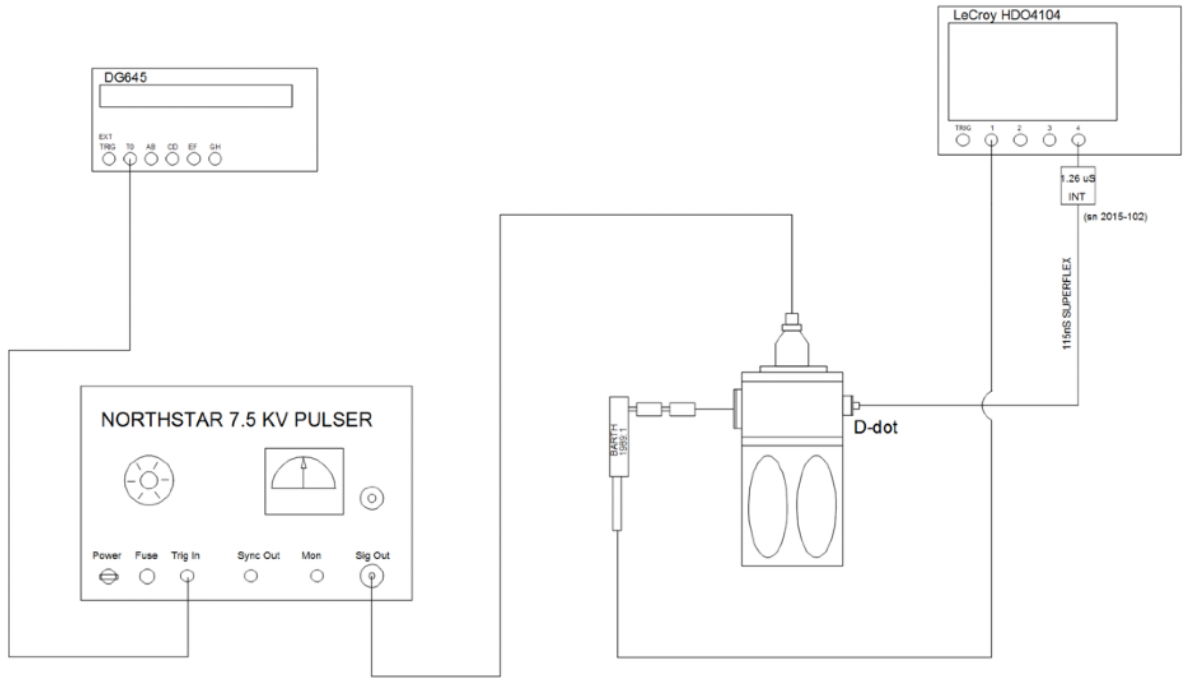

Figure 9.   Recording setup for compensation can D-dot calibration.

Table 3.   Results for D-Dot SN 1E406. Average scale factor (with integrator): $4.975\times10^5$

| D-Dot SN | Shot | dB | Integrator RC [$\mu$s] | Scale Factor |
|---|---|---|---|---|
| 1E406 | 5 | 0 | 1.26 | $4.974\times10^5$ |
| 1E406 | 6 | 0 | 1.26 | $4.957\times10^5$ |
| 1E406 | 7 | 0 | 1.26 | $4.976\times10^5$ |
| 1E406 | 8 | 0 | 1.26 | $4.963\times10^5$ |
| 1E406 | 9 | 0 | 1.26 | $5.004\times10^5$ |

Table 4.   Results for D-Dot SN 1E107. Average scale factor (with integrator): $4.943\times10^5$

| D-Dot SN | Shot | dB | Integrator RC [$\mu$s] | Scale Factor |
|---|---|---|---|---|
| 1E107 | 1 | 0 | 1.26 | $4.948\times10^5$ |
| 1E107 | 2 | 0 | 1.26 | $4.946\times10^5$ |
| 1E107 | 3 | 0 | 1.26 | $4.948\times10^5$ |
| 1E107 | 4 | 0 | 1.26 | $4.947\times10^5$ |
| 1E107 | 5 | 0 | 1.26 | $4.926\times10^5$ |

Table 5.  Results for D-Dot SN 1E395. Average scale factor (with integrator): $4.929 \times 10^5$

| D-Dot SN | Shot | dB | Integrator RC [$\mu$s] | Scale Factor |
|---|---|---|---|---|
| 1E395 | 0 | 0 | 1.26 | $4.909 \times 10^5$ |
| 1E395 | 1 | 0 | 1.26 | $4.947 \times 10^5$ |
| 1E395 | 2 | 0 | 1.26 | $4.921 \times 10^5$ |
| 1E395 | 3 | 0 | 1.26 | $4.933 \times 10^5$ |
| 1E395 | 4 | 0 | 1.26 | $4.935 \times 10^5$ |

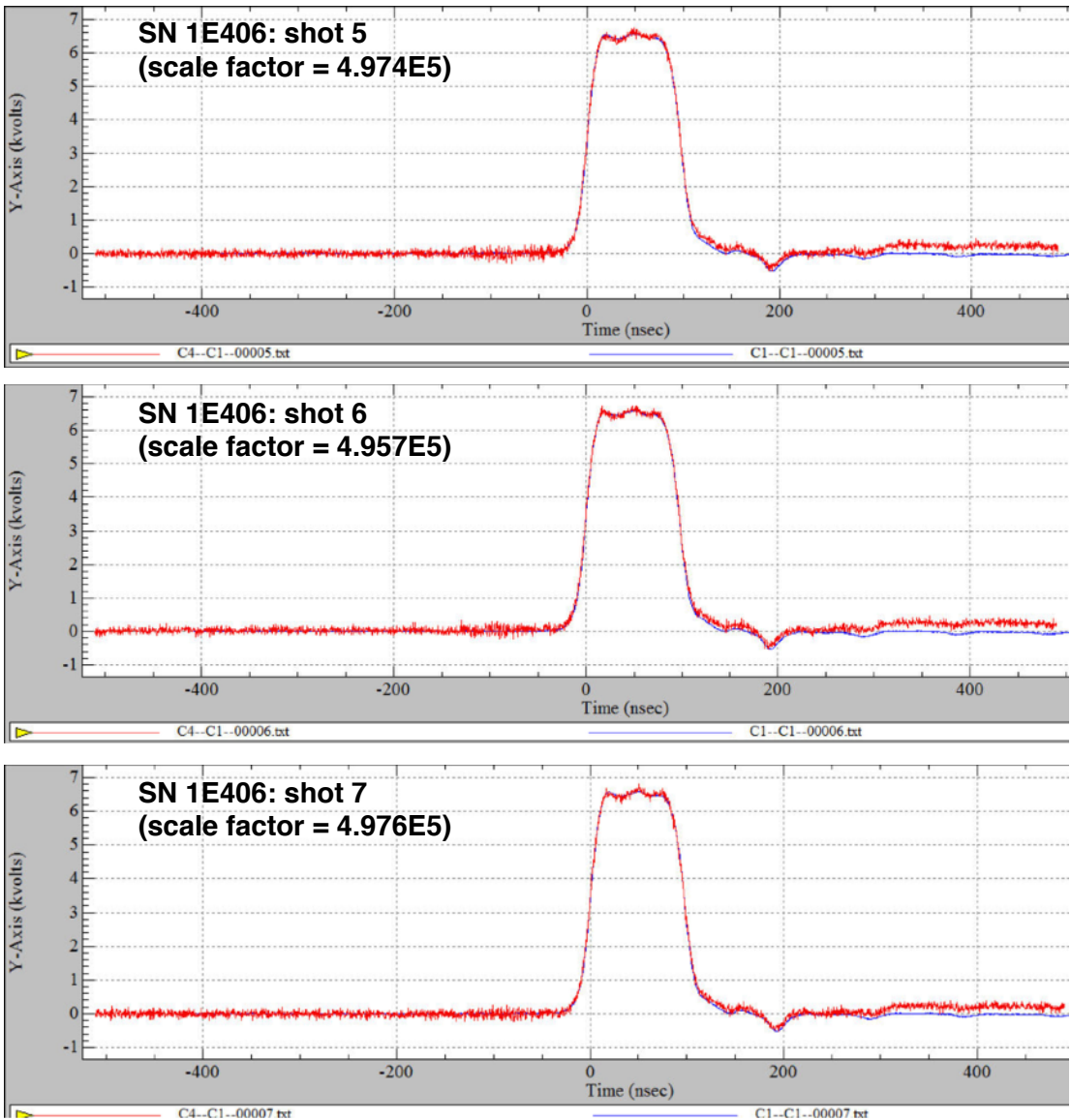

Figure 10.  Calibration waveforms for D-dot SN 1E406, shots 5-7.

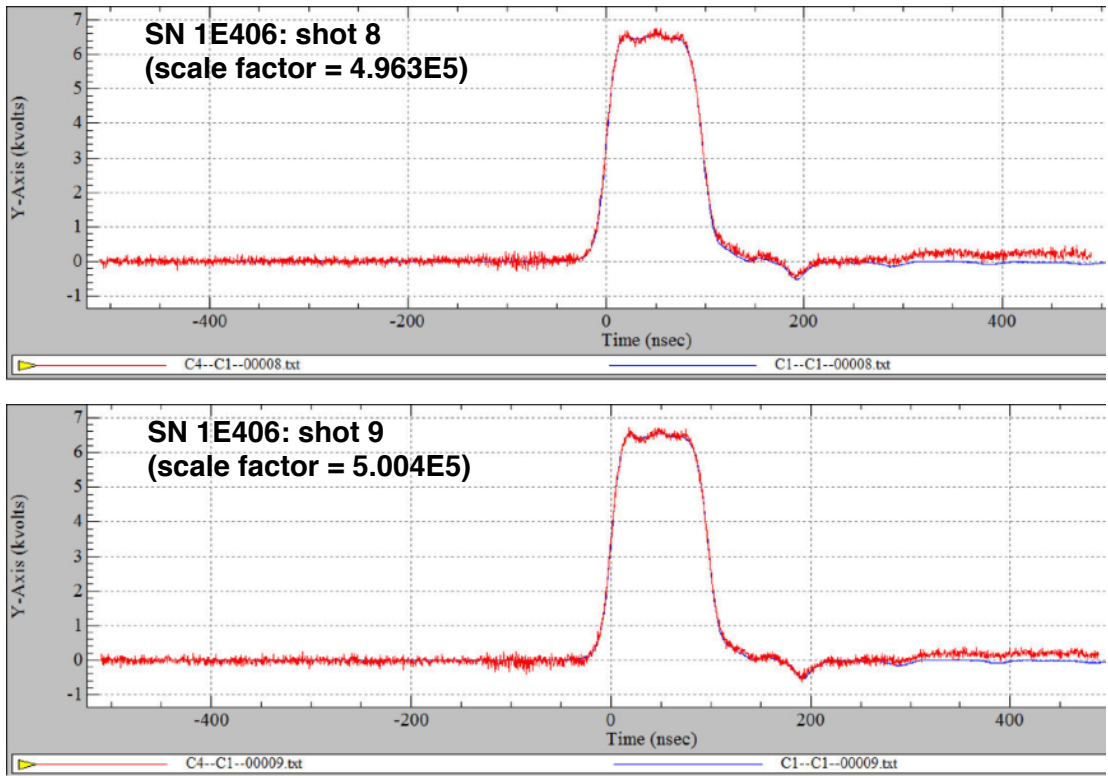

Figure 11. Calibration waveforms for D-dot SN 1E406, shots 8-9.

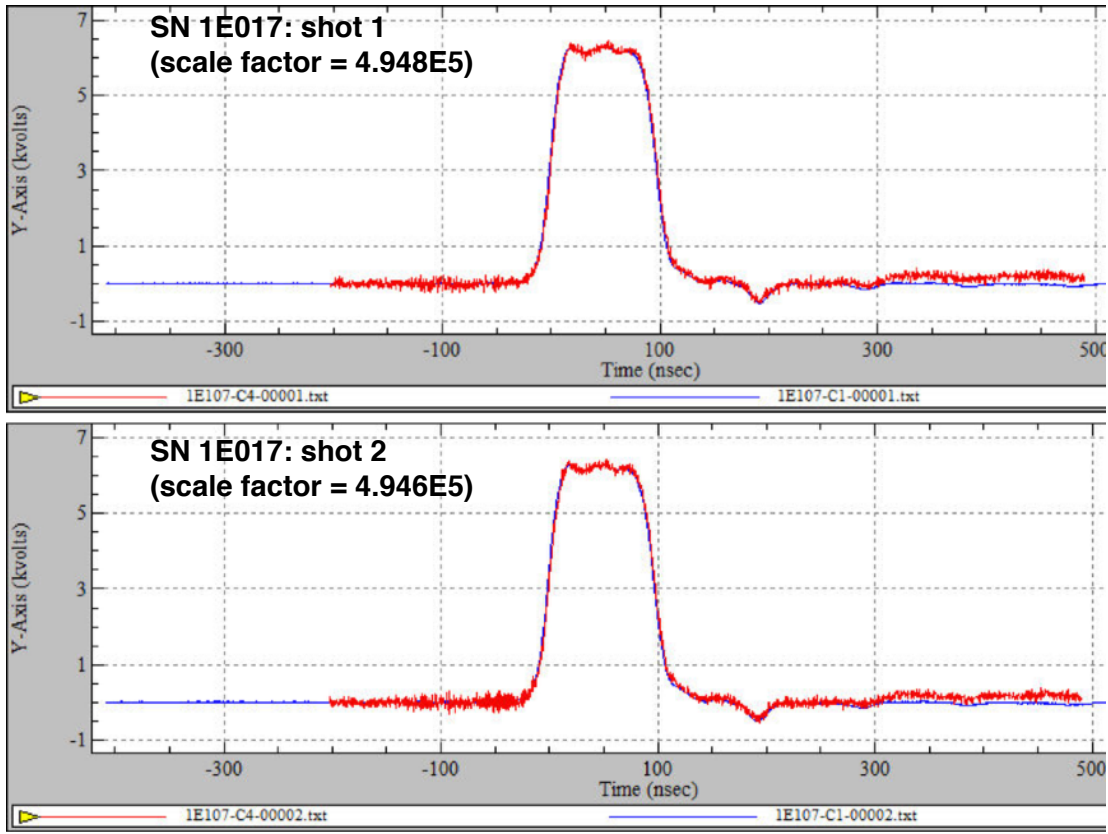

Figure 12. Calibration waveforms for D-dot SN 1E017, shots 1-2.

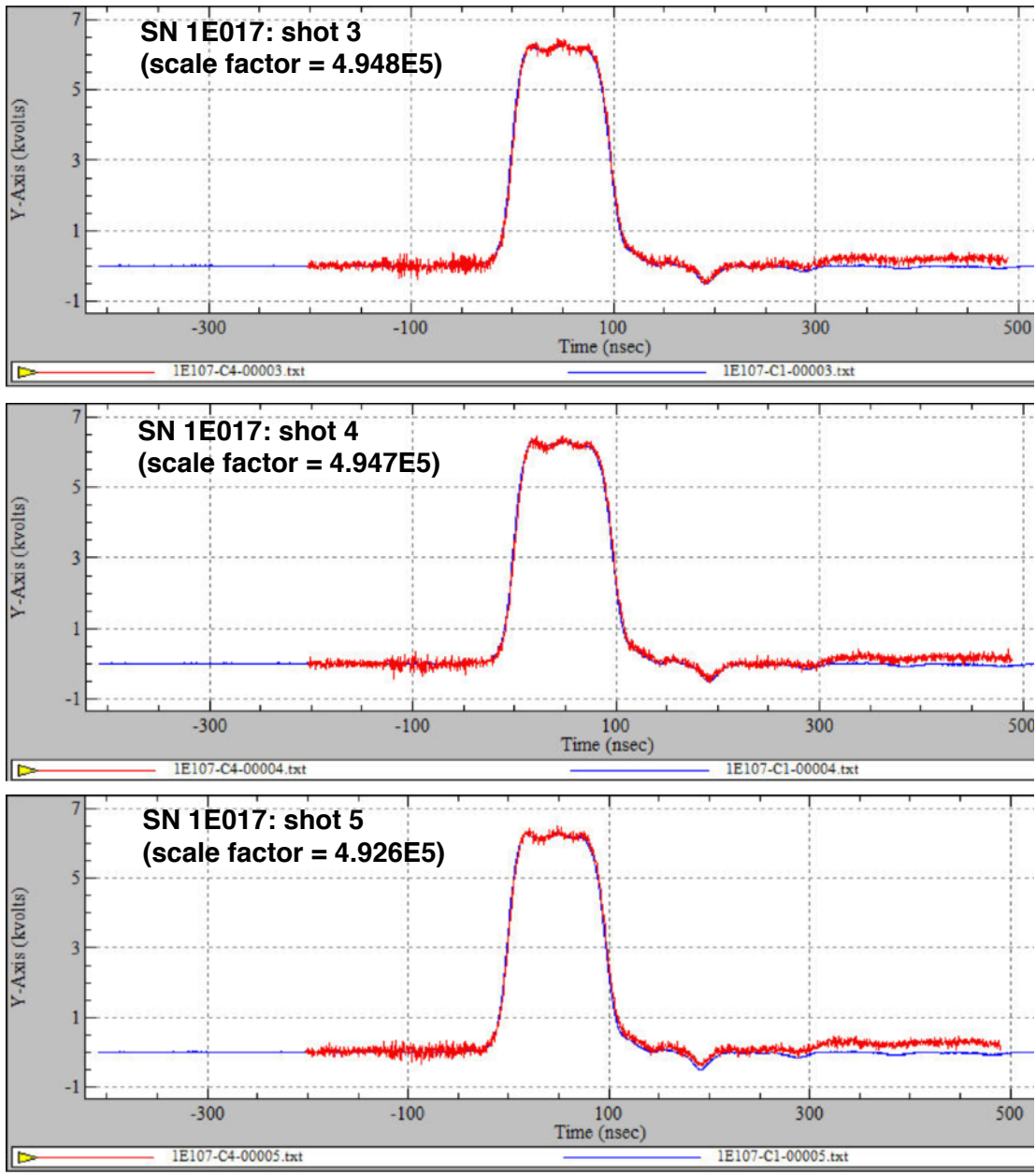

Figure 13. Calibration waveforms for D-dot SN 1E017, shots 3-5.

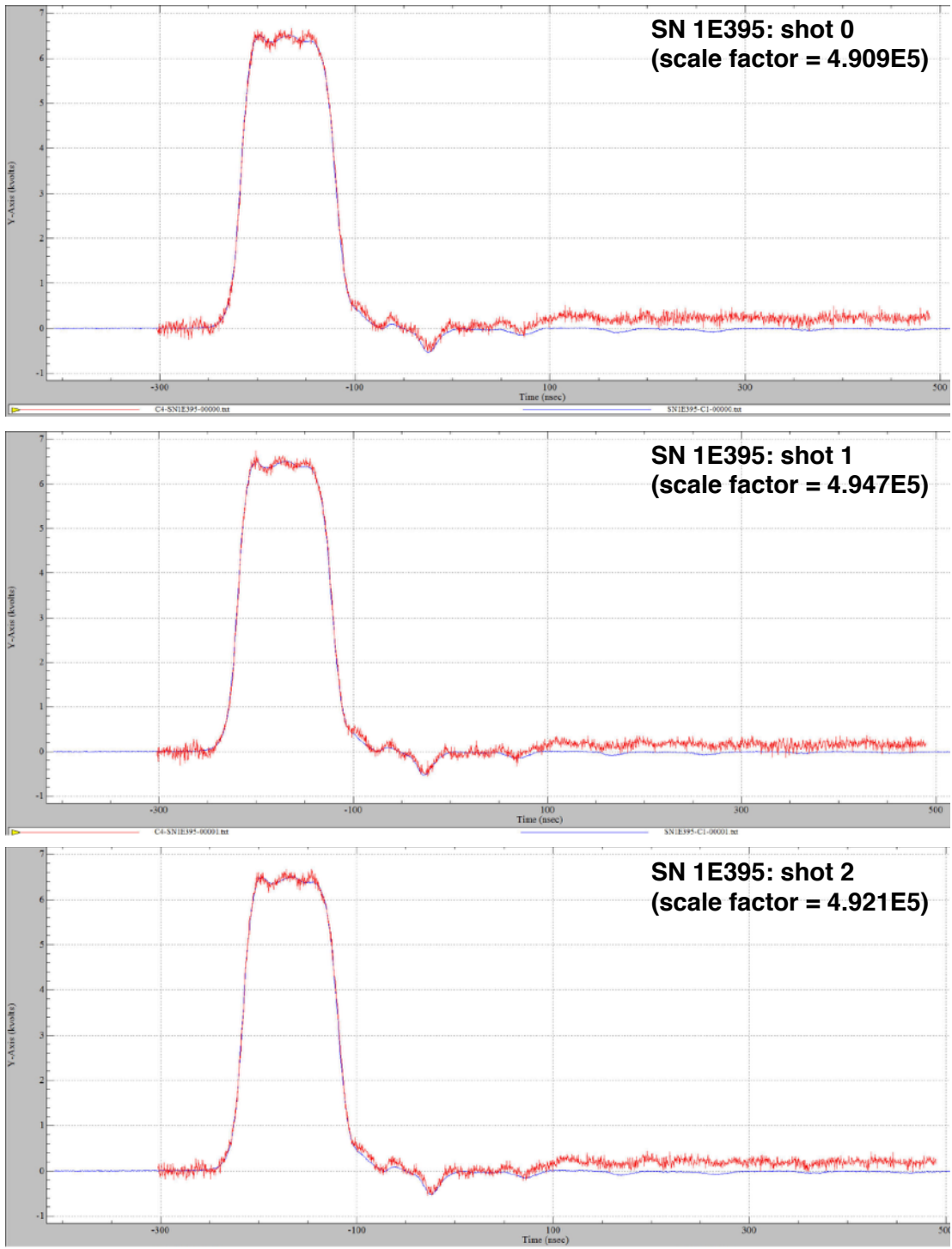

Figure 14. Calibration waveforms for D-dot SN 1E395, shots 0-2.

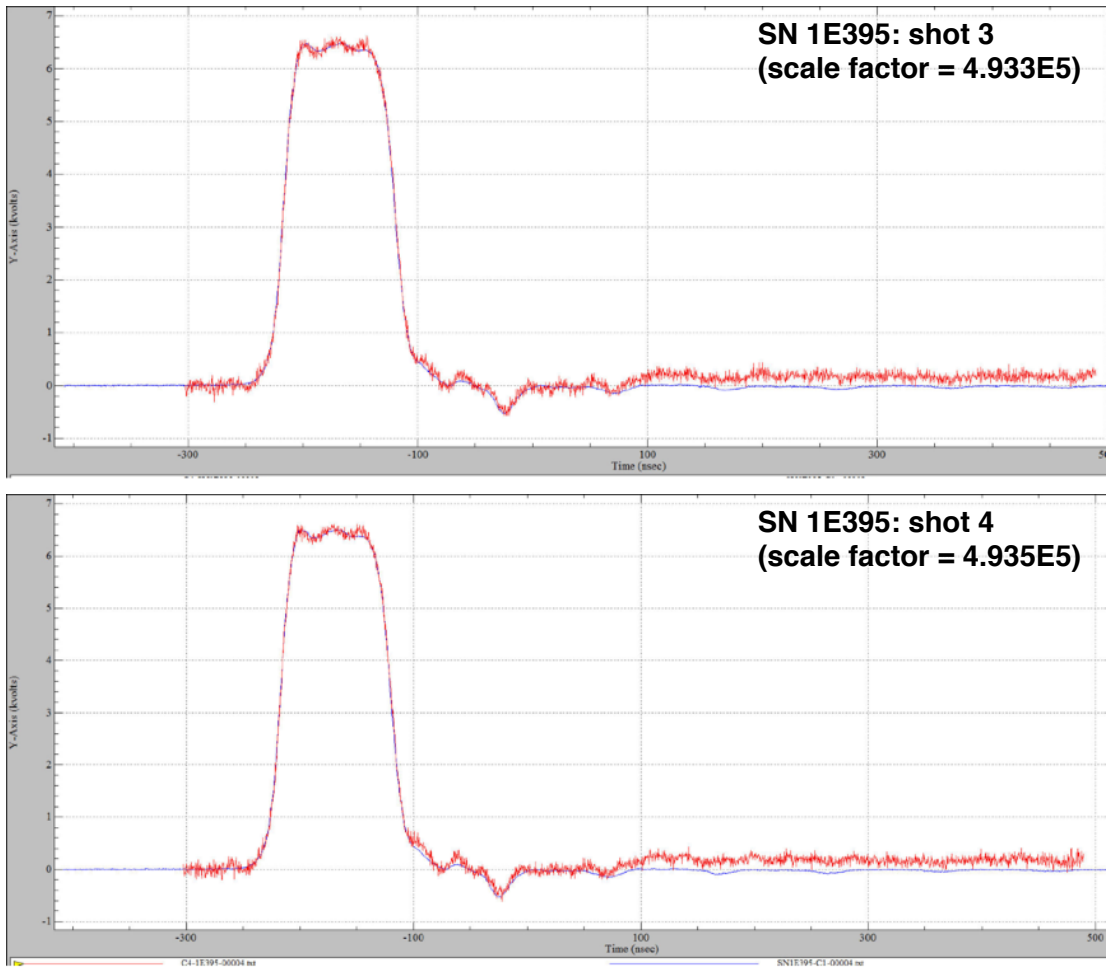

Figure 15. Calibration waveforms for D-dot SN 1E395, shots 3-4.

Assuming all D-dots to be the same, and averaging the scale factors obtained for each one, we obtain that the overall scale factor and capacitance are $(4.949 \pm 0.023) \times 10^5$ and $(101.8 \pm 0.48)$ fF, respectively.

**References**

1   CA Ekdahl, "Voltage and current sensors for a high-density z-pinch experiment," Review of scientific Instruments **51** (12), 1645-1648 (1980).
2   Voss Aerospace, "V-Retainer Coupling — Design Guidelines For Aerospace/Defense Applications".